\documentclass[12pt]{iopart}

\pdfoutput=1

\usepackage{amssymb}
\usepackage{amsbsy}
\usepackage{amsfonts}
\usepackage{graphicx}
\usepackage{curves}

\newcommand{\kt}{\tilde{\kappa}}
\newcommand{\sn}{\mathrm{sn}}
\newcommand{\dn}{\mathrm{dn}}
\newcommand{\cn}{\mathrm{cn}}

\begin{document}

\rightline{BI-TP 2013/19}
\rightline{TUW-13-13}

\title[Critical scalar field collapse in $\mathrm{AdS}_3$]{Critical scalar field collapse in $\mathbf{AdS_3}$:\\an analytic approach}

\author{R Baier$^1$, S A Stricker$^2$ and O Taanila$^1$}

\address{$^1$ Faculty of Physics, University of Bielefeld, D-33615 Bielefeld, Germany}

\address{$^2$ Institute of Theoretical Physics, Technical University of Vienna,\\~~Wiedner Hauptstr.~8-10, A-1040 Vienna, Austria}

\ead{baier@physik.uni-bielefeld.de}
\ead{stricker@hep.itp.tuwien.ac.at}
\ead{olli.taanila@iki.fi}

\begin{abstract}
We present an analytical solution of a massless scalar field collapsing in a three dimensional space-time with a negative cosmological constant, i.e.~asymptotically $\mathrm{AdS}_3$. The Einstein and scalar field equations are formulated using double null Poincar\'e coordinates. Trapping horizons form when a critical parameter is $p > 1$. There are indications that the  horizon radius $r_{AH}$ scales like $r_{AH} \sim (p-1)^{\frac{1}{4}}$ and the black hole mass as $M \sim r^2_{AH} \sim (p-1)^{\frac{1}{2}}$, i.e. with a critical mass exponent of $\frac{1}{2}$.
\end{abstract}

\maketitle

\section{Introduction}

The gravitational collapse of a scalar field in an asymptotically anti-de Sitter space-time can be used as a model of thermalization of a strongly coupled field theory using the AdS/CFT correspondence \cite{Maldacena:1997re,Gubser:1998bc,Witten:1998qj,Son:2007vk}. Recently holographic thermalization has been explored using numerical methods as well
\cite{Garfinkle:2011tc,Wu:2012ri,Wu:2013qi}. Numerous work is devoted to $\mathrm{AdS}_3$ 
\cite{Husain:2000vm,Pretorius:2000yu,Garfinkle:2000br,
Clement:2001bc,
Clement:2001ak,Clement:2001ns}, and issues related to stability
\cite{Bizon:2011gg,Bizon:2013}.

Of particular interest are critical phenomena \cite{Gundlach:2007gc}, which were first observed in scalar field collapse  by Choptuik   \cite{Choptuik:1992jv}  with three important features: scaling, echoing  and universality. The fate of the space-time depends only on one parameter, $p$, which characterizes the initial data of the scalar field.
If the initial data is large enough a black hole forms, and otherwise the scalar field scatters away and flat space-time remains. There is a critical value, $p^*$, that separates the two scenarios and near critically, i.e.~$p\simeq p^*$, the mass of the black hole exhibits a scaling law
\begin{equation}
\label{scaling}
M\sim (p-p^*)^\gamma\;,
\end{equation}
where the critical exponent $\gamma$ is universal with respect to the initial data.
The case of $AdS_3$ was considered in \cite{Pretorius:2000yu}, the five dimensional case in \cite{Bland:2007sg} and the framework for arbitrary dimensions in \cite{Husain:2002nk}
\footnote{For higher dimensional  spherical symmetric collapse without cosmological constant see \cite{Bland:2005kk, Sorkin:2005vz}}. It was conjectured  in \cite{AlvarezGaume:2006dw} that the critical exponent of the scalar field collapse corresponds to the BFKL scaling exponent  $\gamma_{BFKL}$ on the field theory side, supported by the numerical gravity  calculation of \cite{Bland:2007sg} which lies within the 1\% error margin of the field theory calculation. The echoing stems from the self-similarity of the critical solution and can be obtained from the periodicity of the scalar field
\begin{equation}
\label{echo}
\phi(r \rme^{\Delta }, t \rme^{ \Delta })=\phi(r, t),
\end{equation}
where $\Delta$ is the echoing period.

In this paper we derive analytic solutions for the collapse of a massless scalar field in $\mathrm{AdS}_3$ starting from a  scaling ansatz for the dynamical variables
\cite{Wang:2001xba}.
The paper is organized as follows. In section~\ref{sec:equations} we present the Einstein-scalar field equations in double-null
coordinates in the Poincar\'e patch of $\mathrm{AdS}_3$,
and make an ansatz of self-similarity
(in analogy to the discussion of the Roberts solution 
\cite{Roberts:1989sk,Oshiro:1995rp,Soda:1996nq} for the collapse of a scalar field \cite{Brady:1994aq}). 
The novel solutions are given in section~\ref{sec:solutions} and their properties are analyzed in section~\ref{sec:properties}. In the Appendix we review the derivation of the BTZ black hole solution of  \cite{Banados:1992wn,Banados:1992gq} in double null coordinates and also present an alternative derivation of the critical exponent. Finally we end with our conclusions.

\section{Einstein - scalar field equations}
\label{sec:equations}

The line element in the double-null coordinates $u$  and $v$ in the Poincar\'{e} patch  is given by 
\begin{equation}
\rmd s^2 = -\rme^{2 \sigma(u,v)} \rmd u~\rmd v + \frac{r(u,v)^2}{L^2} \rmd x^2
\label{1}
\end{equation}
with the cosmological constant $\Lambda = -\frac{1}{L^2}$.  The Einstein-scalar field equations with a real massless field $\phi(u,v)$ are given by  $(a,b = u,v)$
\begin{equation}
G_{ab} - \frac{g_{ab}}{L^2} = T_{ab} = \partial_a \phi \partial_b \phi -
\frac{1}{2} g_{ab} \partial_c \phi \partial^c \phi,
\label{2}
\end{equation}
and can be rewritten in terms of the Ricci tensor as
\begin{equation}
R_{ab} = - \frac{2}{L^2} g_{ab} + \partial_a \phi \, \partial_b \phi.
\label{3}
\end{equation}
In the following we keep the curvature radius finite and set $L=1$, i.e.~all the dimensionful quantities $r, u ,v$ are scaled by L and thus treated dimensionless. The gravitational constant is also set to $\kappa = 1$. Explicitly the equations of motion for the various components in (\ref{3}) become (with $\partial_u = \frac{\partial}{\partial {u}}$
and $\partial_v = \frac{\partial}{\partial {v}}$)
\begin{eqnarray}\label{eom1}
r (\partial_u \phi)^2 =2 (\partial_u \sigma) (\partial_u r) - \partial^2_u r,\\
r (\partial_v \phi)^2 =2 (\partial_v \sigma) (\partial_v r) - \partial^2_v r,\\
(\partial_u \phi) (\partial_v \phi) = - 2 (\partial_{u,v} \sigma) - \frac{1}{r} (\partial_{u, v} r) - \rme^{2 \sigma},\\
2 (\partial_{u,v} r) = -r\,\rme^{2 \sigma} \; . \label{eom1a}
\end{eqnarray}
The Klein-Gordon equation for the scalar field reads
\begin{equation}
(\partial_u r) (\partial_v \phi) + (\partial_v r) (\partial_u \phi) +
2 r (\partial_{u, v} \phi) = 0.
\label{5}
\end{equation}
In the presence of the scalar field, the Ricci scalar is modified as
\begin{equation}
R = (\partial_a \phi) (\partial^a \phi) - 6~.
\label{A1}
\end{equation}

To solve this system of equations analytically, we make a self-similar ansatz where we introduce the dimensionless coordinate $\eta \equiv \frac{u}{v}$. The scalar field and the metric components in (\ref{1}) are expressed as
\begin{equation}
\phi (u, v) \equiv \Psi (\eta) + \kt \ln v,
\label{6}
\end{equation}
where we set $\kt = 0$ in the following,
\begin{equation}
r(u,v) \equiv v^\alpha f (\eta),~~~\alpha~ =~ {\rm constant},
\label{7}
\end{equation}
and
\begin{equation}
\rme^{2 \sigma(u,v)} = v^{-2} \rme^{2\rho(\eta)} \; .
\label{8}
\end{equation}
This ansatz leads to a system of equations where only the variable $\eta$ appears.
Denoting  $\frac{\rmd}{\rmd \eta} =\,'$ the equations of motion (\ref{eom1}-\ref{eom1a}) can be written as
\begin{eqnarray}\label{eom2}
f\rme^{2 \rho (\eta)}=-2 [(\alpha -1) f^\prime - \eta f^{\prime\prime}],\\
f \Psi^{\prime 2} = 2 \rho^\prime f^\prime - f^{\prime\prime} \label{4b'},\\
\alpha [2 \eta (f^\prime - \rho^\prime f) - (\alpha + 1) f] = 0\label{4c'},\\
-f \eta \Psi^{\prime 2} = 2 (\rho^\prime + \eta \rho^{\prime\prime}) f -
(\alpha -1) f^\prime + \eta f^{\prime\prime}-f \rme^{2 \rho}\;.\label{4d'}
\end{eqnarray} 
The Klein-Gordon equation (\ref{5}) reads
\begin{equation}
[f \Psi^\prime]^\prime = \frac{\alpha -2}{2 \eta} [f \Psi^\prime],
\label{5'}
\end{equation}
which is solved by
\begin{equation}
f \Psi^\prime (\eta) = A \eta^{\frac{\alpha}{2} -1},~~~A = {\rm constant} \; .
\label{9}
\end{equation}

\section{Solutions}\label{sec:solutions}

Despite the non-linear structure of (\ref{eom2}-\ref{4d'}), somewhat surprisingly, analytic solutions can be found. This can be done by recasting the equations of motion into an equation for an anharmonic oscillator as we now show.
The non-linear equation for $f$ is derived from (\ref{4b'}), (\ref{4c'}) and
finally (\ref{9}),
\begin{equation}
\eta f f^{\prime\prime} - 2 \eta f^{\prime 2} + (\alpha + 1)
f f^\prime + A^2 \eta^{\alpha -1} = 0 \; .
\label{33}
\end{equation}
Equivalently a first order equation for $f^\prime$ can be derived,
\begin{equation}
2 \eta^2 f^{\prime 2} - 2 \alpha \eta f f^\prime - A^2 \eta^\alpha -
\frac{\tilde{c}}{2} \eta^{- \alpha} f^4 = 0 \; ,
\label{34}
\end{equation}
which uses the solution of (\ref{4c'}), where $\alpha \not= 0$,
\begin{equation}
\rme^{2 \rho (\eta)} = \frac{\tilde{c} f^2}{\eta^{\alpha+1}}\;.
\label{35}
\end{equation}
In the following it is, however, convenient to continue with (\ref{33}) and
write it as
\begin{equation}
f^3 \frac{d}{d \eta} \bigg[ \frac{\eta f^\prime - \alpha f}{f^2} \bigg] +
A^2 \eta^{\alpha -1} = 0.
\label{36}
\end{equation}
Introducing the new variable $x$ implicitly by defining
\begin{equation}
\eta^\alpha = \left( \frac{2A}{\alpha} \right)^2 \rme^{-2 \sigma x},~~
\sigma = {\rm const.},
\label{37}
\end{equation}
and replacing $f$ by a new function $y(x)$,
\begin{equation}
f = \left( \frac{2A}{\alpha} \right) \eta^{\frac{\alpha}{2}} \frac{1}{y(x)},
\label{38}
\end{equation}
an equation for an ``anharmonic'' oscillator follows \cite{Goenner}
\begin{equation}
\frac{1}{\sigma^2} \frac{d^2}{dx^2} y (x) - y(x) - y^3 (x) = 0 \; .
\label{39}
\end{equation}
This can be integrated to give
\begin{equation}
\frac{1}{2} \left( \frac{dy}{dx} \right)^2 - \frac{\sigma^2}{2} y^2 -
\frac{\sigma^2}{4} y^4 = E = {\rm constant} \; .
\label{40}
\end{equation}
This equation is equivalent to (\ref{34}) by identifying $y$ by $f$ and
fixing the constants by
\begin{equation}
\frac{\tilde{c}}{2} = \frac{\alpha^4}{A^2} \left( \frac{E}{4 \sigma^2} \right).
\label{41}
\end{equation}
Without loss of generality the constant $A$ is fixed by $A = \frac{\alpha}{2}$,
and for $\sqrt{1 - \frac{4E}{\sigma^2}} =p$, the following cases have to be
distinguished,
\begin{equation}
\begin{array}{rll}
{\rm (i)~~}& p = 0 \; , & \frac{4E}{\sigma^2} = 1 \\
{\rm (ii)~~}& 0< p \le 1 \; , & 0 \le \frac{4E}{\sigma^2} < 1\\
{\rm (iii)~~}& p > 1  \; ,& \frac{4E}{\sigma^2} < 0 \; .
\end{array}
\label{42} 
\end{equation}
Equation (\ref{40}) is integrated in terms of the elliptic integral
\begin{equation}
\frac{1}{\sqrt{2}} \ln \bigg[ \eta^{- \frac{\alpha}{2}} \bigg] =
\frac{\alpha}{2 \sqrt{2}} \ln \left( \frac{1}{\eta} \right) = \int^\infty_y
\frac{\rmd y^\prime}{\sqrt{ (y^{\prime 2} + p + 1) (y^{\prime 2} -p + 1)}} \; ,
\label{43}
\end{equation}
with the boundary at $\eta = \frac{u}{v} = 1, y = \infty$ i.e.~$r = 0$.
It is important to distinguish the cases (i) - (iii) of (\ref{42}).

\subsection*{(i) $\mathbf{p=0:}$}

For $p = 0$ the integral of (\ref{43}) can be performed to give
\begin{equation}
r (u,v) = (uv)^{\frac{\alpha}{2}} \cot \bigg[ \frac{\alpha}{2 \sqrt{2}}
\ln \frac{v}{u} \bigg] \; .
\label{44}
\end{equation}
This is very similar to the solution (\ref{15}).

\subsection*{(ii) $\mathbf{0 < p \le 1:}$}

For $0 < p < 1$ the integral of (\ref{43}) can be given in terms of the Jacobian elliptic functions $\sn (\hat{u}|m)$ and $\cn (\hat{u}|m)$,
\begin{equation}
r(u,v) = \frac{(uv)^{\frac{\alpha}{2}}}{\sqrt{1+p}}
\frac{\sn \left( \sqrt{\frac{1+p}{8}} \alpha \ln \frac{v}{u} \bigg\vert
\frac{2p}{1+p} \right)}{\cn \left( \sqrt{\frac{1+p}{8}} \alpha
\ln \frac{v}{u} \bigg\vert \frac{2p}{1+p} \right)}\;,
\label{45}
\end{equation}
where the conventions from  \cite{Stegun} are  used.
An independent solution with the boundary located at $\eta = u/v = 1$ and 
$y = 0$, i.e.~$r \rightarrow  \infty$, reads
\begin{equation}
r(u,v) = \frac{(uv)^{\frac{\alpha}{2}}}{\sqrt{1-p}}
\frac{\cn \left( \sqrt{\frac{1+p}{8}} \alpha \ln \frac{v}{u} \bigg\vert
\frac{2p}{1+p} \right)}{\sn \left( \sqrt{\frac{1+p}{8}} \alpha
\ln \frac{v}{u} \bigg\vert \frac{2p}{1+p} \right)} \; .
\label{455}
\end{equation}
In the special case $p=1$, or $(m=1)$, this reduces to
\begin{equation}
r(u,v) = \frac{1}{2 \sqrt{2}} \big[ v^\alpha - u^\alpha \big].
\label{46}
\end{equation}
All the above solutions do not have an apparent horizon, since
$g^{ab} \,\partial_a r \, \partial_b r \not=0$. 

\subsection*{(iii) $\mathbf{p>1:}$}

For the case  $p > 1$ the solution is again given in terms of the Jacobi elliptic functions 
\begin{equation}
r(u,v) = v^\alpha f (\eta) = \frac{(uv)^{\frac{\alpha}{2}}}{\sqrt{2p}}
\frac{\sn \left( \sqrt{\frac{p}{4}} \alpha \ln \frac{v}{u} \bigg\vert 
\frac{1+p}{2p} \right)}{\dn \left( \sqrt{\frac{p}{4}} \alpha \ln
\frac{v}{u} \bigg\vert \frac{1+p}{2p} \right)}\;,
\label{47}
\end{equation}
with 
\begin{equation}
\dn^2 (\hat u\vert m) = 1 - m~ \sn^2 (\hat u \vert m)\; , ~~ m = \frac{(1+p)}{2p} \; ,~~\hat{u}= \sqrt{\frac{p}{4}} \alpha \ln \frac{v}{u}  \; .
\label{48}
\end{equation}
Since $p>1$, $m$ is limited to be $1 > m > \frac{1}{2}$.

For $p > 1$ another independent solution for $r$ with the boundary $\eta = 1$ 
at $y = \sqrt{p - 1}$ is given by
\begin{equation}
r(u,v)  = \frac{(uv)^{\frac{\alpha}{2}}}{\sqrt{p - 1}}~
\cn \left( \sqrt{\frac{p}{4}} \alpha \ln \frac{v}{u} \Bigg\vert 
\frac{1+p}{2p} \right) ~,
\label{477}
\end{equation}
which is periodic in $\hat{u}$ as the solution (\ref{47}).

\section{Properties of the solutions}

\label{sec:properties}

We are now ready to investigate the properties of the solutions. For all the cases  considered above, the metric (\ref{1}) can be written as
\begin{equation}
ds^2 = - \frac{\alpha^2}{2} (1 - p^2) \frac{r^2}{(uv)^\alpha } \frac{du dv}{uv}
 + r^2 dx^2 \; ,
\label{488}
\end{equation}
where we have used $\tilde{c} = \alpha^2 (1 - p^2)/2$. The metric now depends only on the different radial functions $r(u,v)$ obtained in the previous section.
For compact notation, we suppress the arguments of the function $r(u,v)=r$, but it should be remembered that $r$ is not a coordinate but rather a function of the coordinates $u$ and $v$.

The Ricci scalar  (\ref{A1}) becomes  
\begin{equation}
R = - \frac{2}{(p^2 - 1)} \frac{(uv)^{2 \alpha}}{r^4}  - 6 \; .
\label{A2}
\end{equation} 
In the limit $ r \rightarrow 0$ this is singular but approaches the value of pure AdS if the first term on the right hand side vanishes.

To investigate the presence of an apparent horizon, we compute the expression\cite{Poisson:1990eh}
\begin{equation}
g^{ab} \partial_a r \partial_b r = 2 g^{ u  v} \partial_{ u} r
\partial_{ v} r = r^2 - M \; ,
\label{22}
\end{equation}
which vanishes for $r^2_{AH} = M > 0$ (at the apparent horizon), where
$M$ is the Misner-Sharp black hole mass \cite{Misner}.
Therefore an apparent horizon is only present if a solution to  
\begin{equation}
g^{ab} \partial_a r \partial_b r \vert_{r = r_{AH}} = 0
\label{49}
\end{equation}
exists.
One can check that the family of solutions for which $0<p\le 1$ never satisfies the above condition and a naked singularity forms at $r(u,v)=0$ as can be seen from (\ref{A2}).
Therefore, in the following we focus on the solutions with $p>1$ which show critical behaviour and the formation of an apparent horizon.

\subsection{Black hole formation for $p>1$}

In this section we investigate the properties of the space-time of (\ref{47}). 
In order to make this  solution more transparent it is convenient
 to start from the metric of (\ref{488}) and introduce the new variables $U$ and $V$ by the transformations
\begin{equation}
U = \frac{p-1}{\sqrt{2}} \frac{1}{u^\alpha} > 0  \qquad \mathrm{and} \qquad V =- \frac{p+1}{\sqrt{2}}\frac{1}{v^\alpha} < 0.
\end{equation}
%\be
%(uv)^\alpha = \frac{p^2-1}{2}\left( \frac{1}{-UV} \right) \; .
%\ee
This leads to the conformally flat metric
\begin{equation}\label{dsUV}
ds^2 = r^2(U,V) \left[ -dU\,dV+dx^2\right]\;,
\end{equation}
where
\begin{equation}
\label{r}
r = \frac{\sqrt{p^2-1}}{2\sqrt{p}\sqrt{-UV}}
\frac{\sn(\hat{u}|m)}{\dn(\hat{u}|m)} \; ,
\end{equation}
where
\begin{equation}
\hat{u} = \frac{\sqrt{p}}{2}\ln \left[ \frac{p+1}{p-1}\left( \frac{U}{-V} \right)\right] \; \mathrm{and} \quad m = \frac{p+1}{2p} \; .
\end{equation}
The function $r$ is periodic because of the properties of the elliptic function  $\sn (\hat u|m)$,
\begin{equation}
\label{period}
sn(\hat{u}=2n K(m)|m)=0 \; ,
\end{equation}
where the period is given by
\begin{equation}
2 K(m) = \int^{\frac{\pi}{2}}_0 d \theta [1 - m~\sin^2 \theta]^{-\frac{1}{2}} \; , \quad n=1,2,\ldots
\end{equation}
In what follows we will concentrate on the first interval where $r$ is positive 
\begin{equation}
\label{interval}
\frac{p-1}{p+1} \leq \left( \frac{U}{-V}\right) \leq \frac{p-1}{p+1} \exp \left[ \frac{4K(m)}{\sqrt{p}}\right] \; .
\end{equation}
All the conclusions are valid for the other intervals as well, since they are merely repetitions of the first interval. Patched together they fill out the whole space-time. This is a consequence of  the self-similarity imposed by the ansatz we made for the metric functions $r$ and $\sigma$ in (\ref{1}). 

With the further transformation
\begin{equation}
U = z + t \;, \quad -V = z-t\;,
\end{equation}
the metric becomes diagonal, $\quad -\rmd U\,\rmd V = -\rmd t^2 + \rmd z^2$.
In the limit $z \gg t$, $-U/V \simeq 1 + \mathcal{O}(t/z)$, the metric approaches the $\mathrm{AdS}_3$-vacuum one,
\begin{equation}
ds^2 \to \frac{\hat{L}^2}{z^2} \left( -dt^2 + dz^2 + dx^2 \right)
\end{equation}
where the curvature is given by $\hat{L}^2 \to 1/4$ for $p\gg1$.\\

\subsection*{Apparent horizon}

To identify the location of the apparent horizon we have to find solutions to (\ref{49}) which reads
\begin{equation}
g^{ab}\,\partial_a r\,\partial_b r =  \frac{2 (p^2-1)}{\alpha UV} 
\frac{m \bigg[ \sn^4 (\hat u|m) - 2 p ~\sn^2 (\hat u|m) + \frac{2p^2}{(1+p)}
\bigg] }{\sn^2 (\hat u|m)\dn^2 (\hat u|m)}=0.
\label{51}
\end{equation}
The expression in the square brackets can be factorized as
\begin{equation}
\left[ \sn^2 (\hat u|m) - p \left( 1 - \sqrt{\frac{p-1}{p+1}} \right) \right]
\left[ \sn^2 (\hat u|m) - p \left( 1 + \sqrt{\frac{p-1}{p+1}} \right) \right] \; .
\label{52}
\end{equation}
When the expression in the first square brackets vanishes, we get
\begin{equation}
\sn^2 \left( \sqrt{\frac{p}{4}} \alpha \ln \frac{1}{\eta} \Bigg|
\frac{1+p}{2p} \right) \Bigg|_{AH} = p \left( 1 - \sqrt{\frac{p-1}{p+1}}
\right) \le 1,
\label{53}
\end{equation}
indicating the presence of an apparent horizon.

Using the concept of trapping horizons introduced in \cite{Hayward:1994bu}, we can find the properties of the apparent horizon by looking at the derivatives  ($U\neq0, V\neq0$)
\begin{eqnarray}
\partial_U r = \frac{\partial r}{\partial U } = 
-\frac{\sqrt{p^2-1}}{4\sqrt{p}}
 \frac{1}{(-UV)^{1/2}(U)} \left[ \frac{\sn(\hat{u}|m)}{\dn(\hat{u}|m)}-\sqrt{p}\frac{\cn(\hat{u}|m)}{\dn^2(\hat{u}|m)}\right]\;,\\
\partial_V r =  \frac{\partial r}{\partial V }
 = \frac{\sqrt{p^2-1}}{4\sqrt{p}} \frac{1}{(-UV)^{1/2}(-V)}
 \left[ \frac{\sn(\hat{u}|m)}{\dn(\hat{u}|m)}+\sqrt{p}\frac{\cn(\hat{u}|m)}{\dn^2(\hat{u}|m)}\right] \;,\\
 \partial_{U,V}r = \frac{\partial^2 r}{\partial U\,\partial V}
= -\frac{(p^2-1)^{3/2}}{16p^{3/2}(-UV)^{3/2}}
 \frac{\sn^3(\hat{u}|m)}{\dn^3(\hat{u}|m)} \; .
 \end{eqnarray}
When $\partial_V r = 0$ it follows from (\ref{53}) that
\begin{equation}
 \sn(\hat{u}|m) = -\sqrt{p}\, \frac{\cn(\hat{u}_{AH}|m)}{\dn(\hat{u}_{AH}|m)} = \left[ p \left( 1-\sqrt{\frac{p-1}{p+1}}\right)\right]^\frac{1}{2} < 1 
\end{equation}
at the apparent horizon. Therefore $\partial_U r <0$ and $\partial_{U,V}  r < 0$ at  $r_{AH}$,  which implies the presence of a "future  outer trapping horizon" (FOTH), i.e.~a black hole  at $\hat{u} = \hat{u}_{AH}$, according to the definitions of \cite{Hayward:1994bu}.

\subsection*{Critical exponent}

Knowing the precise location of the apparent horizon enables us to determine the critical exponent from the expression of the mass (\ref{22}) which takes the compact form 
\begin{equation}
\label{M}
M = \frac{p}{(-UV)} \frac{\dn^2(\hat{u}|m)}{\sn^2(\hat{u}|m)}=\frac{p^2-1}{4(-UV)^2r^2}
\end{equation}
  and  is positive everywhere. At the apparent horizon the mass takes the simple form
\begin{equation}
 M_{\mathrm{AH}}=r_{AH}^2=\frac{\sqrt{p^2-1}}{2(-UV)},
\end{equation}
for fixed $UV$. 
Near the critical value $p \sim 1$ the mass scales with the exponent $\gamma = 1/2$ as defined in (\ref{scaling}). In order to eliminate the divergence for $UV \to 0$, which is due to the breakdown of the self-similarity, one may normalize by the mass (\ref{M})  (see the discussion in \cite{Soda:1996nq,Oshiro:1994hd}), 
by keeping the ratio $\sn/\dn = \mathrm{constant} = 1$,
\begin{equation}
M_{norm} = \frac{p}{-UV}\;,
\end{equation}
such that 
\begin{equation}
\label{Mcrit}
\frac{M_{AH}}{M_{norm}} = \frac{\sqrt{p^2-1}}{2 p }\sim (p-1)^\frac{1}{2}.
\end{equation}
In \ref{A3} the critical exponent is obtained following the lines  of \cite{Soda:1996nq, Koike:1995jm}. This method does not need normalization and uses the fact that there is a fixed saddle point that bifurcates  the phase flow between the black hole and the flat space-time.
Using this method we also find a  critical mass exponent of $\gamma=\frac{1}{2}$

In \cite{Husain:2000vm} V. Husain and M. Olivier calculate numerically an exponent of
$\gamma = 1.62$ and Choptuik and Pretorius  \cite{Pretorius:2000yu} find $\gamma/2=1.2\pm 0.05$ using a different method.
% The different values of the exponents depend on the solutions \cite{Maison:1995cc}.

\subsection*{Scalar field}

Next we investigate the behaviour of the scalar field. The expression for the scalar field is obtained from  (\ref{9})
\begin{equation}
 \eta \Psi^\prime (\eta) = 
\frac{A}{\eta^{ - \frac{\alpha}{2}}}
\frac{1}{f} = \alpha\sqrt{\frac{p}{2}}  \mathrm{ds} 
\left(\hat u |m \right),
\label{57}
\end{equation}
in terms of the elliptic function $\mathrm{ds} = \dn/\sn$ and has poles at $\sn
(\hat u \vert m) = 0$, i.e.~for $\hat u =  2 n  K(m)$.
This equation can be integrated and thus the scalar field is (up to an integration constant) given by
\begin{equation}
\phi(u,v) = \Psi(\hat{u}) = 
\frac{1}{\sqrt{2} } \ln \frac{1 - \cn(\hat u  \vert m)}
{1 + \cn(\hat u  \vert m)} ~,
\label{psi}
\end{equation}
with $\Psi(\hat{u}) = 0$ at $\hat{u} =  K(m =\frac{1+p}{2p} )$
and diverging at $\hat{u} = 0$.

At the past singularity, $r\to 0$ for $\hat{u}=0$, the field is diverging like $\phi\to -\infty$, since $\cn(\hat{u}|m)\to 1$.
In the asymptotic AdS region, $-U/V\to 1$, the scalar field is constant.
%By adding a suitable constant of integration such that $\phi_\mathrm{AdS}=0$,  pure AdS is obtained.
At the horizon $\hat{u}_{AH}$  the field is finite and increasing, $\phi_{AH}>\phi_\mathrm{AdS}$. Finally, as $r\to0$, $-U/V \to \frac{p-1}{p+1}\exp\left[ \frac{4K(m)}{\sqrt{p}}\right]$ (cf.~(\ref{interval})), and the field is diverging like
\begin{equation}
\phi \stackrel{r\to0}{\longrightarrow} \mathrm{const} \times |\ln r| \to \infty \; .
 \end{equation}
The space-time starts in a singularity, forms an apparent horizon and finally  ends in a big crunch.  Starting at $\phi_\mathrm{AdS}$ it crosses the horizon $\phi_{AH}$ and diverges (implodes) approaching the singularity $r\to 0$.
Therefore we can replace the past region $t<0$ by pure $\mathrm{AdS}_3$ and match it smoothly to the $t>0$ region. Once the scalar field is turned on it is impossible to reach the asymptotic boundary and everything will cross the horizon and end in the singularity.

\begin{figure*}
\centering
\includegraphics[width=7.8cm]{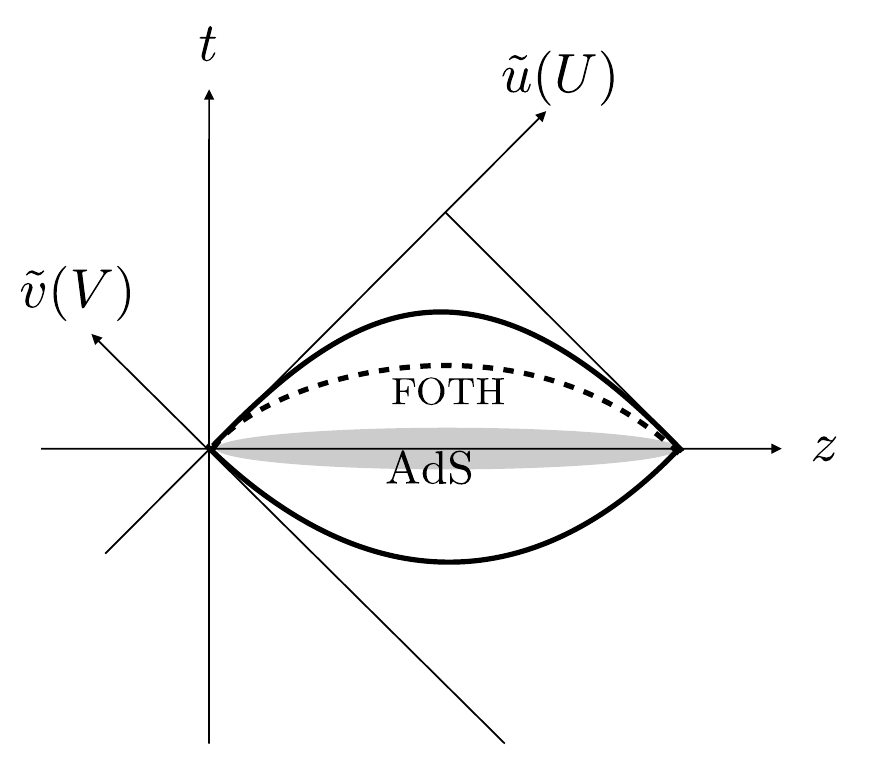}\includegraphics[width=5cm]{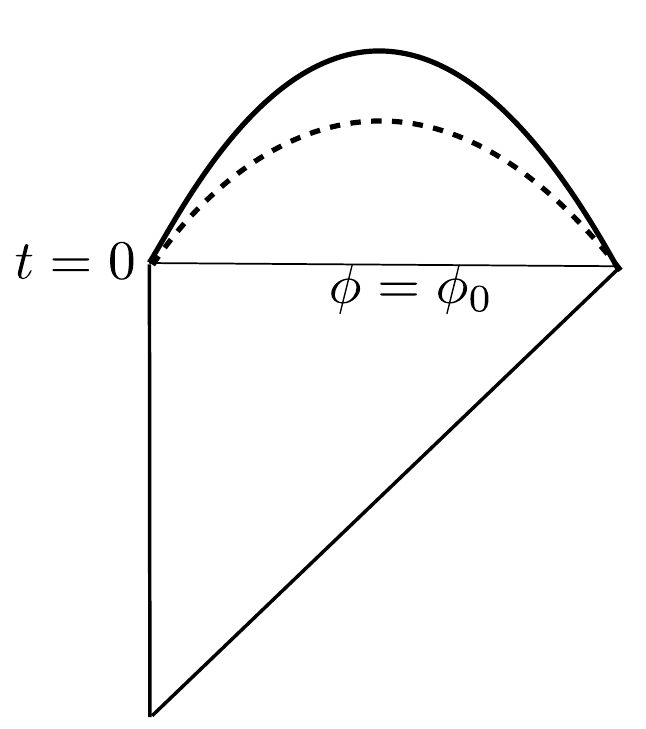}
\caption {On the lefthand side is the Penrose diagram for the space-time describing the scalar field collapse. The solid curves are the past and future singularities and the dashed line is the future outer trapping horizon (FOTH). The grey area corresponds to empty AdS space. On the righthand side is the Penrose diagram of a space-time that is constructed by taking the upper half of the space-time describing the scalar field collapse and gluing it to the lower part of an empty $\mathrm{AdS}_3$. This corresponds to a space-time which starts as empty $\mathrm{AdS}$, but where a scalar field is then turned on at $t=0$, which then collapses.}
\label{Penrose}
\end{figure*}

Due to the periodicity of the elliptic functions, we can read off the echoing period (\ref{echo}) and is simply given by
\begin{equation}
\Delta=2 K(m)\;,
\end{equation}
with $K(m)$ given in (\ref{period}).
In summary, the Penrose diagram in figure~\ref{Penrose} shows the behaviour of the solution (\ref{47}), by using the coordinates $U = \tan \tilde{u}$ and $V=-\tan \tilde{v}$.
For the solution (\ref{477}), the situation is essentially the same.

%\setlength{\unitlength}{0.8cm}
%
%\begin{figure*}
%\centering
%\begin{picture}(8,8)
%\put(0,4){\vector(1,0){8}}
%\put(1,0){\vector(0,1){8}}
%\put(0,3){\vector(1,1){5}}
%\put(5,0){\vector(-1,1){5}}
%\put(4,7){\line(1,-1){3}}
%\put(4,1){\line(1,1){3}}
%\thicklines
%\put(1,4){\bezier{0}(0,0)(3,-2)(6,0)}
%\thinlines
%\curvedashes[1mm]{1,0.5,0.5,0.5}
%\put(1,4){\bezier{0}(0,0)(3,3)(6,0)}
%\curvedashes[1mm]{1,1}
%\put(1,4){\bezier{0}(0,0)(3,-3)(6,0)}
%\put(-0.5,5.2){$U(\bar{u})$}
%\put(4.5,8.2){$V(\bar{v})$}
%\put(0.8,8.2){$T$}
%\put(8.2,3.8){$X$}
%\put(6.9,4.2){\small$\mathrm{AdS}$}
%\put(3.5,5.6){\small$r=0$}
%\put(3.5,2.1){\small{$r=0$}}
%\put(3.4,3.2){\small{$\mathrm{POTH}$}}
%\put(7,4){\circle*{0.3}}
%\end{picture}
%\end{figure*}

\section{Conclusion}

In this paper we analyzed the collapse of a massles scalar field in 3 dimensional space-time with a negative cosmological constant. After imposing self-similarity on the fields we find novel analytic continuous self-similar solutions in terms of Jacobi elliptic functions. In accordance with previous studies we find that the fate of the space-time depends only on one parameter. The solutions show critical behaviour with a scaling law for the mass in the  critical region 
\begin{equation}
M_{AH}\sim (p-1)^\frac{1}{2}\;,
\end{equation}
with a critical exponent $\gamma=\frac{1}{2}$. 
If the parameter $p$ is large enough, $p>1$, an apparent horizon forms.
For these solutions the space-time emerges from an initial singularity, forms an apparent horizon and ends in a big crunch and repeats itself for eternity.
 If  the initial condition is in the range $0<p<1$ the scalar field collapses to form a naked singularity. This is expected since the scalar field has no potential which would counterbalance the gravitational force.
A critical exponent of $\gamma=\frac{1}{2}$ was also found for self-similar  spherical scalar field collapse in four dimensions \cite{Soda:1996nq, Oshiro:1994hd}, in the case of the collapse of thin rings of dust \cite{Peleg:1994wx} and in the two-dimensional RST model \cite{Strominger:1993tt}.
 Interestingly, in \cite{Emparan:2013xia} it was conjectured that $\gamma=\frac{1}{2}$ is the exact solution for collapse in the limit of infinite dimensions.
 
\ack

We thank  D.~Grumiller, V.~Ker\"anen, G.~Kunstatter, L.~Thorlacius and  A.~Vuorinen for useful comments.
S.~S.~was supported   by the START project Y435-N16 of the Austrian Science
Fund (FWF) and O.~T.~by the  Sofja Kovalevskaja program of the Alexander von Humboldt foundation. 

\appendix

\section{Case \boldmath{$\alpha = 0$} and vanishing field  \boldmath{$\phi = 0$}}

In order to have reference solutions to the set of equations (\ref{eom2}) we show how the pure AdS and BTZ metric are obtained in double null coordinates  for  the case of of a vanishing field $\phi = \Psi = 0$. 
Furthermore, in eq. (\ref{4c'}) two cases have to be distinguished, namely
$\alpha = 0$ and $\alpha \not= 0$.
For $\alpha = 0$ we have, noting that $r = f$, 
\begin{equation}
2 \eta (r^\prime - \rho^\prime r) -r = \mathrm{constant} \equiv M,
\label{10}
\end{equation}
and
\begin{equation}
2 \rho^\prime r^\prime = r^{\prime\prime}.
\label{11}
\end{equation}
Introducing the variable $\xi = \ln \eta$, (\ref{10}) and (\ref{11}) can
be combined to
\begin{equation}
r\,\partial^2_\xi r - 2 (\partial_\xi r)^2 + M (\partial_\xi r) = 0,
\label{12}
\end{equation}
which is solved by
\begin{equation}
\partial_\xi r = \frac{M}{2} -  \frac{r^2}{2},
\label{13}
\end{equation}
where an integration constant is chosen to be $-1$. 
%(The
%coefficient $\rme^{2 \sigma} = -\frac{2}{uv} \frac{\partial^2_\xi r}{r}$).
Furthermore (with $r=0$ for $u = v$)
\begin{equation}
\ln \frac{u}{v} = \xi = 2 \int^r_0 \frac{dr^\prime}{M-r^{\prime 2}} =
\left\{ \begin{array}{l@{\,,}l}
- \frac{2}{\sqrt{-M}} \arctan \frac{r}{\sqrt{-M}} & ~~ M<0 \\
\frac{2}{\sqrt{M}} \mathrm{arctanh} \frac{r}{\sqrt{M}} & ~~ M>0 
\end{array} \right. \; .
\label{14ab}
\end{equation}
For the case $M< 0$ we choose $M = -1$, i.e.
\begin{equation}
r = \tan \left(\frac{1}{2} \ln \frac{v}{u}\right)
\label{15}
\end{equation}
and with (\ref{8}) and (\ref{11}) 
\begin{eqnarray}
\rme^{2 \sigma} & = \frac{1}{v^2} \rme^{2 \rho} = \frac{1}{v^2} \exp
\bigg[- \xi - 2 \ln \cos \left( \frac{\xi}{2} \right) \bigg] \nonumber \\
& = \frac{1}{uv} \frac{1}{\cos^2 \left( \frac{\xi}{2} \right)} =
\frac{1}{uv} (r^2 + 1)
\label{16}
\end{eqnarray}
With the transformation $u=\rme^{\hat u}$ and $v=\rme^{\hat v}$, the metric
reads $(L=1)$
\begin{equation}
\rmd s^2 = - (r^2+1) \rmd \hat u \, \rmd \hat v + r^2 \rmd x^2,
\label{17}
\end{equation}
$r = \tan \left( \frac{\hat v - \hat u}{2} \right)$. Furthermore, with the
transformation (see (\ref{14ab}))
\begin{eqnarray}
\hat u & = t - \arctan r, \nonumber \\
\hat v & = t + \arctan r,
\label{18}
\end{eqnarray}
the vacuum $\mathrm{AdS}_3$ space is obtained,
\begin{equation}
\rmd s^2 = - (r^2+1) \rmd t^2 + \frac{\rmd r^2}{r^2+1} + r^2 \rmd x^2.
\label{19}
\end{equation}
The BTZ black hole solution, the case $M>0$, reads from (\ref{14ab})
for $r \le \sqrt{M}$,
\begin{eqnarray}
r & = \sqrt{M} \tanh \left( \sqrt{M} \frac{\xi}{2} \right), \nonumber \\
& \equiv \sqrt{M} \tanh \left( \sqrt{M} \frac{(\hat u - \hat v)}{2} \right),
\label{20}
\end{eqnarray}
with the metric
\begin{eqnarray}
\rmd s^2 & = \left( 1 - \frac{r^2}{M} \right) M \frac{\rmd u\,\rmd v}{uv} + r^2\rmd x^2 \\
& = M \left( 1- \frac{r^2}{M} \right) \rmd \hat u\,\rmd \hat v + r^2 \rmd x^2 \nonumber \\
& = - (-M+r^2) \rmd t^2 + \frac{\rmd r^2}{-M+r^2} + r^2 \rmd x^2.
\label{21}
\end{eqnarray}
The presence of the black hole is also seen, when the
expression (\ref{22}) is calculated \cite{Poisson:1990eh},
%\be
%g^{ab} \partial_a r \partial_b r = + 2 g^{\hat u \hat v} \partial_{\hat u} r
%\partial_{\hat v} r = r^2 - M,
%\label{22}
%\ee
which vanishes for $r^2_{AH} = M > 0$ (at the apparent horizon), and
$M$ is the (square) of the BTZ black hole mass 
\cite{Banados:1992wn,Banados:1992gq}.

\section{Case {\boldmath ${\alpha \not= 0}$}  and 
{\boldmath ${\phi = \Psi = 0}$}}

Starting from (\ref{4b'}) and (\ref{4c'}) the case $\alpha \not= 0$
(but $\Psi = 0$) becomes
\begin{equation}
\rme^{2 \sigma} = \frac{\rme^{2 \rho}}{v^2} = \frac{\tilde{c} f^2}{v^2 
\eta^{\alpha + 1}},~~~\tilde{c} = {\rm const.}
\label{23}
\end{equation}
with the equation for $f$ being
\begin{equation}
\eta f f^{\prime\prime} - 2 \eta f^{\prime 2} + (\alpha+1) f f^\prime = 0 \; .
\label{24}
\end{equation}
This can be recast to the form
\begin{equation}
f^3 \frac{d}{d \eta} \bigg[ \frac{\eta f^\prime - \alpha f}{f^2} \bigg]
= 0 \; .
\label{25}
\end{equation}
Requiring that the expression in the aquare brackets is constant, we get
\begin{equation}
\eta f^\prime - \alpha f + p^2 f^2 = 0,~~p^2 = {\rm const.},
\label{26}
\end{equation}
which can be integrated by
\begin{equation}
\int \frac{\rmd \eta}{\eta} = \ln \eta = \int \frac{\rmd f}{\alpha f - p^2 f^2}
= \frac{1}{\alpha} \ln \frac{f}{\alpha - p^2 f} \; .
\label{27}
\end{equation}
The expression for $r$ is then gfiven by
\begin{equation}
r = v^\alpha f = \frac{\alpha u^\alpha}{1 + p^2 \eta^\alpha} =
\frac{(uv)^\alpha}{c_1 u^\alpha + c_2 v^\alpha},
\label{28}
\end{equation}
with $c_1 = \frac{p^2}{\alpha},~c_2 = \frac{1}{\alpha}$.
The solution looks simpler if one makes the transformation $\hat u = u^{-\alpha}$ and $\hat v =
v^{-\alpha}$, giving
\begin{equation}
\rmd s^2 = - \frac{\tilde c}{\alpha^2} r^2 \rmd \hat u \, \rmd \hat v + r^2 \rmd x^2 \; , \quad r = \frac{1}{c_1 \hat v + c_2 \hat u} \; .
\label{29}
\end{equation}
Choosing e.g.~$\frac{\tilde c}{\alpha^2} = 1$, and transforming
\begin{equation}
\rmd \hat u = \rmd t - \frac{\rmd r}{r^2}\quad \mathrm{and} \quad \rmd \hat v = \rmd t + \frac{\rmd r}{r^2}
\label{31}
\end{equation}
gives then the $\mathrm{AdS}_3$ vacuum metric for $r \gg 1$.
\begin{equation}
\rmd s^2 = -r^2 \rmd t^2 + \frac{\rmd r^2}{r^2} + r^2 \rmd x^2 \; .
\label{32}
\end{equation}

\section{Critical exponent from an autonomous system}\label{A3}

For an alternative derivation of the critical exponent we start with the
Eddington-Finkelstein metric,
\begin{equation}
ds^2 = - \bar g g~du^2 - 2g~du~dr+ r^2 dx^2,
\label{58}
\end{equation}
with two dimensionless functions $\bar g (u,r)$ and $g (u,r)$
\cite{Brady:1994aq}. Furthermore the curvature radius  of AdS is set to unity.
Self-similarity is introduced by using the dimensionless coordinate $\eta = \frac{r}{-u} =
\frac{r}{|u|}$ and by making the ansatz
\begin{equation}
\bar g = \bar g (\eta)\;,\quad g(u,r) = r^{-2} \tilde g (\eta)\;,\quad \phi (u,r) = \Psi (\eta) \; .
\end{equation}
Next it is convenient to  introduce the variable $\xi = \ln \eta$ and the ratios
\begin{equation}
y = \frac{\bar g}{\tilde g} = y (\xi) \quad {\rm and} \quad z = \frac{\eta}{\bar
  g},
\label{61}
\end{equation}
for which the equations of motion take the compact form  % (c.f. \cite{Brady:1994aq}),
\begin{eqnarray}
\dot{y} & = \frac{d}{d \xi} y = 4 \frac{1-z}{1-2z} (1-y), \nonumber \\
\dot{z} & = \frac{y-2}{y} z,\\
\ddot{\Psi} & = \frac{yz-2}{(1-2z)y} \dot{\Psi} \nonumber
\label{62}
\end{eqnarray}
where
\begin{equation}
\dot{\Psi}^2 = \frac{2 (y-1)}{(1-2z)y} \; .
\label{63}
\end{equation}
There is a stationary point at $y = 1, z = 0$ and $\dot{\Psi} = 0$. Using
$\frac{\rmd y}{\rmd z} = \frac{\dot{y}}{\dot{z}}$ one derives a solution in terms
of $h(y) \equiv \left(c \sqrt{1-y} \right) /y$ (for $y < 1,~c = {\rm const.}$)
\begin{equation}
z_{\mp} = - h \bigg[ 1 \mp \sqrt{1 + \frac{1}{h}} 
\bigg].
\label{64}
\end{equation}
Again, an apparent horizon is given by 
\begin{equation}
 g^{ab} \, \partial_a r \, \partial_b r = \frac{\bar g}{g} =
-r^2 y,
\label{65}
\end{equation}
i.e. for $y \to 0$. 
Following \cite{Soda:1996nq} critical behaviour and exponents are obtained by using the linear expansion around the stationary (fixed) point, i.e.
\begin{eqnarray}
\delta \dot{y} & = - 4 \delta y \; ,\\
\delta \dot{z} & = - \delta z \; . \nonumber
\label{66}
\end{eqnarray}
This set of equations has  eigenvalues $\kappa = -1$ and $-4$.
Using the scaling relation 
\begin{equation}
M \sim r^2_{AH} \sim \epsilon^{- \frac{2}{\kappa}}, \epsilon \to 0,
\label{67}
\end{equation}
i.e. with $\epsilon \equiv (p - p^{*}) = (p-1)$ (c.f. (\ref{Mcrit})), one finds for the dominant relevant mode
\begin{equation}
M \sim (p-1)^{\frac{1}{2}}
\label{68}
\end{equation}
with a critical mass exponent of $\gamma=\frac{1}{2}$.

\section*{References}

\bibliographystyle{unsrt} 

\bibliography{refs}

\end{document}